\RequirePackage{fix-cm}
\documentclass[twocolumn]{svjour3}

\smartqed  

\usepackage{graphicx}
\usepackage{amssymb}
\usepackage{caption}
\usepackage{subcaption}
\usepackage{hyperref}
\usepackage{todonotes}
\usepackage{booktabs}
\usepackage{multirow}

\newlength{\twosubht}
\newsavebox{\twosubbox}

\newenvironment{conflicts}{\begin{conflict}}
{\end{conflict}}

\journalname{Computing and Software for Big Science}

\begin{document}

\title{Generating and refining particle detector simulations using the Wasserstein distance in adversarial networks}

\author{Martin Erdmann \and Lukas Geiger \and Jonas Glombitza \and David Schmidt}

\institute{\at
              Physics Institute 3A, RWTH Aachen University, 52056 Aachen, Germany \\
              Tel.: +49-241-80-27330\\
              Fax:  +49-241-80-22189\\
              \email{erdmann@physik.rwth-aachen.de} 
}

\date{Received: date / Accepted: date}

\maketitle

\begin{abstract}
We use adversarial network architectures together with the Wasserstein distance to generate
or refine simulated detector data.
The data reflect two-dimensional projections of spatially distributed signal patterns
with a broad spectrum of applications.
As an example, we use an observatory to detect cosmic ray-induced air showers 
with a ground-based array of particle detectors.
First we investigate a method of generating detector patterns with variable signal strengths
while constraining the primary particle energy.
We then present a technique to refine simulated time traces of detectors 
to match corresponding data distributions.
With this method we demonstrate that training a deep network with refined data-like 
signal traces leads to a more precise energy reconstruction of data events compared to 
training with the originally simulated traces.
\end{abstract}

\section{Introduction}
\label{intro}

Modern deep learning methods have been shown to
be highly successful, e.g., in applications of handwriting, speech, and image recognition
\cite{Hinton,Ciresan,Yu,ILSVRC2012,ResNet1,Go}.

A new training concept is realized in so-called generative adversarial networks (GANs) which produce artificial images from random input while guided by real images \cite{2014arXiv1406.2661G}.
They are based on two networks, an image generator and a discriminator
separating artificial from real images, trained in opposition to one another.
Similarly, adversarial training methods have been used to modify artificial images to better
reproduce patterns found in natural images \cite{2016arXiv161207828S}.

Two improvements of GAN methods which influence this work have recently been reported.
So-called auxiliary classifier generative adversarial networks (AC-GANs) generate artificial images bound to given image class labels using label conditioning \cite{2016arXiv161009585O}.
In addition to quantifying differences between real and artificial images, the
Wasserstein distance has been introduced in generative adversarial networks (WGANs) to improve the stability of the learning process and to avoid mode-collapsing problems which are widely known for other GAN setups \cite{2017arXiv170107875A,2017arXiv170400028G}.
For a review of deep learning methods see \cite{review}.

In both particle and astroparticle physics research, deep learning concepts have already 
been applied successfully for data analyses, see e.g. 
\cite{NoVA,exotics,higgs,kaggle,jetflavor,Baldi:2016fql,ttH,Erdmann:2017str}.
Applications of the GAN concept have demonstrated generation of jet
kinematics and calorimeter showers with unprecedented speed
\cite{deOliveira:2017pjk,Paganini:2017hrr,Paganini:2017dwg}.
In addition, adversarial training methods have been shown to protect classifier networks 
from an error-prone variable \cite{2017PhRvD..96g4034S}.

In this paper we investigate adversarial training with the Wasserstein distance for a number of particle
physics applications.
First we present a method for generating two-di\-men\-sional signal patterns in spatially distributed 
sensors for a given physics label.
This is a general task with broad applications in both particle and astroparticle detector
simulations.

As an example we use cosmic ray-induced air showers in the Earth's atmosphere
which produce signals in ground-based detector stations.
This setup corresponds to a calorimeter experiment with a single readout layer.
We will train a WGAN to generate signal patterns corresponding to a given primary
particle energy.

In a further step, we tackle a pressing matter arising in training deep networks with simulated
data that differ from measured distributions.
We refine simulated signal traces to approximate real data (which, for simplicity, are 
simulated in this paper as well) using an adversarial training concept guided by WGANs.
We then compare the quality of reconstructing particle energies using a deep neural network
after training with either the original or the refined simulated data.

Our paper is structured as follows:
We introduce the Wasserstein distance and explain its application in adversarial training
before presenting our network architectures for generating data or refining simulated data.
After that, we specify the simulated data sets used for training and evaluating the networks.
We then generate data-like signal patterns constrained by energy labels.
We also refine simulated time traces and evaluate their impact on network training
before presenting our conclusions.

\section{Adversarial network architectures}
\label{sec:1}
In the adversarial training method, a generator network is required to learn probability
distributions underlying observed event data distributions.
A discriminator network is used to support this learning process by quantifying the differences 
between a set of event data distributions and the generated event distributions.

In contrast to {\em supervised} machine learning, where network training is performed using 
a true label (e.g. classification as signal or background), adversarial training has no such true label.
Instead, it is based on a similarity measure between two probability distributions and
is thus considered {\em unsupervised} learning.

The feedback of the discriminator network to the generator network about the quality of the
generated events is encoded in the loss function.
When using cross entropy type loss functions, training GANs has been observed to be delicate, hard to monitor and sometimes produce
incoherent results.
A frequently observed issue is a phenomenon known as mode collapsing, where the generator produces results in a restricted 
phase space only.

In the following sections, we first introduce the Was\-ser\-stein distance as an alternative loss 
function in generative adversarial training which leads to improved training stability. Additionally, mode collapsing has not been observed when using the Wasserstein distance \cite{2017arXiv170400028G}.
We then expand the WGAN concept to generating events according to a given label, and to
refining simulated event distributions.

\subsection{Adversarial training with the Wasserstein distance}
\label{sec:1a}

An alternative loss function for adversarial networks has been formulated based on the Wasserstein-1 distance, also referred to as Earth mover's distance \cite{2017arXiv170107875A}.
As an intuitive interpretation, this distance gives the cost expectation for 
moving a probability distribution onto a target distribution along optimal transport paths.

The Wasserstein distance exhibits wanted properties concerning convergence of sequences of 
probability distributions \cite{2017arXiv170107875A}.
This distance measure can thus serve well for quantifying the similarity between data $x$ 
and generated events $\tilde{x}$.
Here, $x$ represents a set of event observables in data, while $\tilde{x}$ represents the corresponding
observables for generated events. 
A common approach for generating $\tilde{x}$ is to implement a neural network $g_{\theta}$ 
with weights $\theta$:
\begin{equation}
    \tilde{x}=g_{\theta}(z)
    \label{eq:g_events}
\end{equation}
Here, $z$ is a random input which can be sampled from an arbitrary distribution.

As computing the formal definition of the Wasserstein distance ($D_{W}$) is intractable, an equivalent representation via the Kantorovich-Rubin\-stein duality is used \cite{KR}:
\begin{equation}
    D_{W} = \sup_{f \in \mathrm{Lip}_1} \left(\mathbb{E}[ f(x) ] - \mathbb{E}[ f(\tilde{x}) ]\right)
    \label{eq:DW}
\end{equation}
Here, $\mathrm{Lip}_1$ is the set of 1-Lipschitz functions $f$, and $\mathbb{E}$ represents
expectation values for $f$ operating on the data events $x$ and the generated events 
$\tilde x$, respectively.

The core idea of the WGAN concept is to approximate the set of 1-Lipschitz functions $f$ using a neural network $f \approx f_w$ parameterized by the weights $w$.
The difference of the expectation values in eq. (\ref{eq:DW})
\begin{equation}
    C_1 = \mathbb{E}[ f_w(x) ] - \mathbb{E}[ f_w(\tilde{x}) ]
    \label{eq:C1}
\end{equation}
constitutes the central term in the loss function of the network.
The network is labeled `critic' as the converged value of $C_1$
gives a measure of the similarity of generated and data events.

Approximating the supremum in (\ref{eq:DW}) is done by minimizing the loss of the critic 
network with the negative argument $-C_1$.
In order to include the Lipschitz condition, the loss function has been
extended by a gradient penalty \cite{2017arXiv170400028G}:
\begin{equation}
    C_2 = \lambda \; \mathbb{E}[(\vert\vert \nabla_{\hat{u}} f_w(\hat{u}) \vert\vert_2 - 1 ) ^2 ]\;.
    \label{eq:C2}    
\end{equation}
Here, the event admixture 
\begin{equation}
    \hat{u} = \varepsilon x + (1-\varepsilon) \tilde{x}
\end{equation}
of real data $x$ and generated data $\tilde{x}$ is used to calculate the gradients of the 
critic network which are forced by the loss to remain close to one. 
The randomly and uniformly drawn value $0\leq \varepsilon \leq 1$ samples the gradients along connecting lines between $x$ and $\tilde{x}$.
$\lambda$ represents a hyperparameter of the training.

The generator uses the gradient of the distance measure $C_1$ (\ref{eq:C1}) with respect to the parameters $\theta$ for training.
In order to provide this measure we first update the critic by
subjecting $m$ data events and $m$ generated events to the network 
represented by $f_w$.

In this initial training step, the weights $w$ of the critic network are optimized to minimize 
the loss $-C_1 + C_2$ from eqs. (\ref{eq:C1}, \ref{eq:C2}).
During this step, the parameters $\theta$ of the generator are frozen.
In the adjacent training step, the critic weights $w$ are frozen temporarily, and the parameters
$\theta$ of the generator network $g_\theta$ are adjusted.
By reducing the Wasserstein distance measure, which is based on the output of the critic 
network, the generator $g_\theta(z)$ is trained to generate more realistic data samples.
The critic is then trained again and the algorithm starts from the beginning.

This iterative procedure is repeated until overall convergence is achieved, leaving
$C_1$ (\ref{eq:C1}) as our measure of similarity between generated and data events.
To provide an accurate gradient for the generator, the critic is usually trained for $n_{\mathrm{critic}}>1$ iterations before updating the generator once.

\begin{figure}
\sbox\twosubbox{
  \resizebox{\dimexpr\linewidth}{!}{
    \includegraphics[height=3cm]{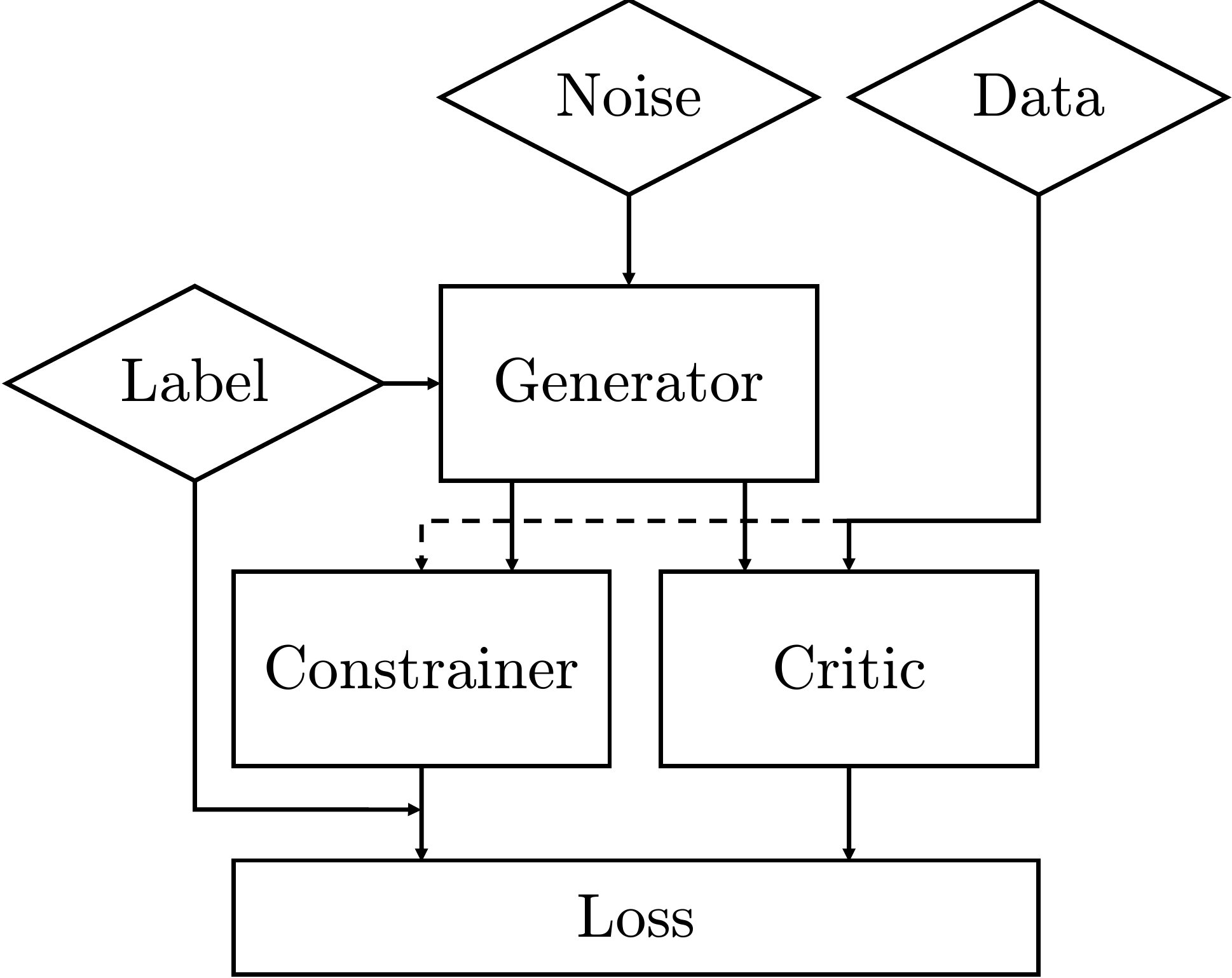}
    \includegraphics[height=3cm]{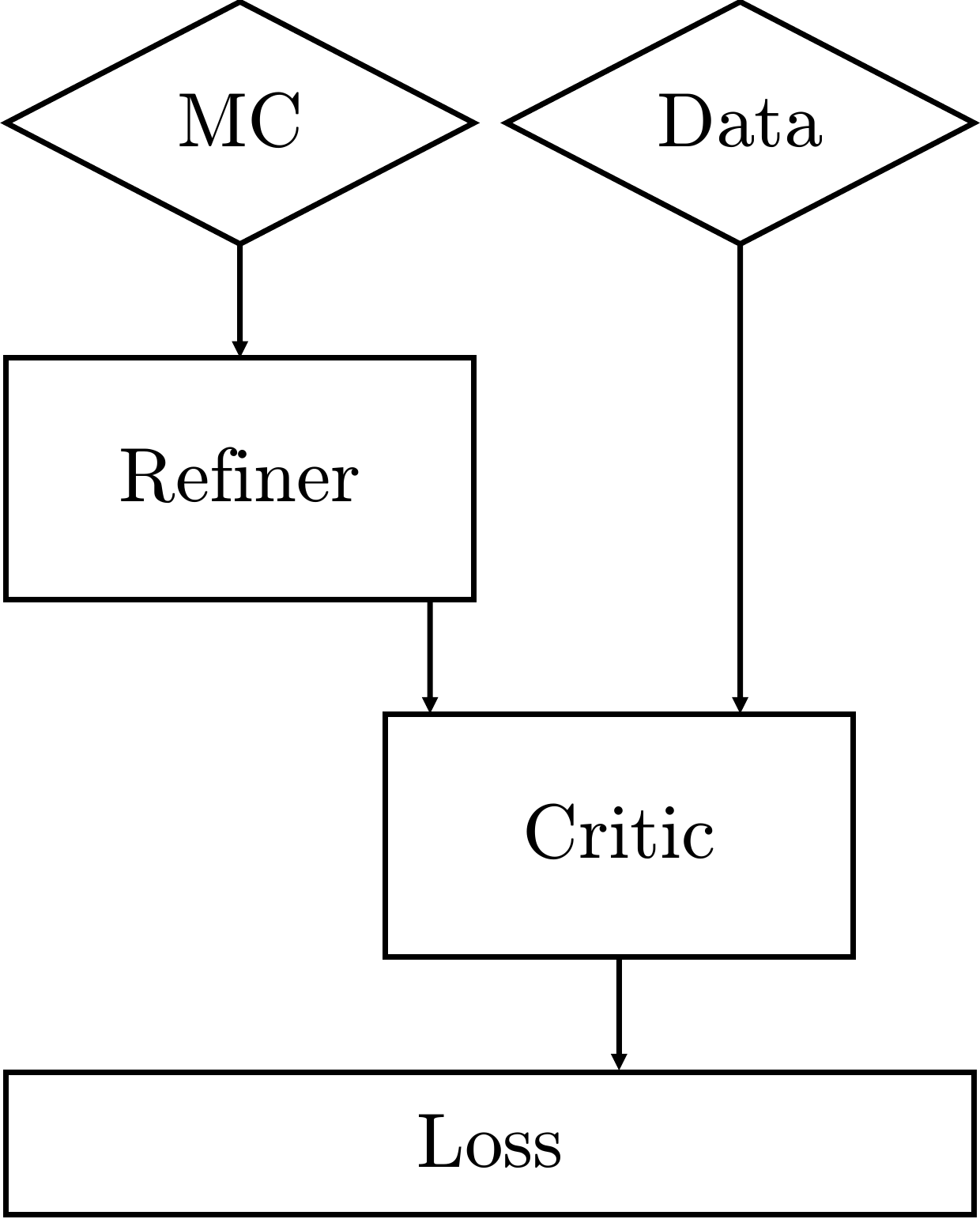}
  }
}
\setlength{\twosubht}{\ht\twosubbox}

\subcaptionbox{\label{fig:architecture_gan}}{
  \includegraphics[height=0.975\twosubht]{network_gan.pdf}
}\hfill
\subcaptionbox{\label{fig:architecture_refiner}}{
  \includegraphics[height=0.975\twosubht]{network_refiner.pdf}
}
\caption{Architectures of 
(a) the generator step of a conditioned generative adversarial network (solid line), 
and supervised training step of the constrainer network (dashed line),
(b) the refining adversarial network.}
\label{fig:architecture}
\end{figure}

\subsection{Physics conditioning of the generator}
\label{sec:1b}

To enforce generated events to reflect certain properties of data events, the input to the generator can 
be extended by physics labels $y_{label}$ in addition to the random numbers $z$ in eq. (\ref{eq:g_events}):
\begin{equation}
    \tilde{x} = g_{\theta}(z, y_{label})
\end{equation}
The required labels can, for example, be particle kinematics where the labels are obtained from corresponding 
energy or angular distributions. 
To push the generator network training towards a label condition, an additional term is introduced in the 
loss function.
It compares the value of an input label with the result 
of an additional network $a_{\theta^\prime}$ that reconstructs the corresponding observable from the generated data:
\begin{equation}
    C_{3} = \kappa\; {[\,y_{label} - a_{\theta^\prime}(\tilde x)\,]^2}
    \label{eq:C3}    
\end{equation}
We will call this network $a_{\theta^\prime}$ `constrainer' network parameterized by the weights $\theta^\prime$.
The weight of the physics label in the total loss function $-C_1 + C_2 + C_3$
of the generator network is controlled by the hyperparameter $\kappa$.
When using several physics labels in a conditioning process, 
the loss function can be extended accordingly.

The constrainer network $a_{\theta\prime}$ is trained supervised using data and their 
associated physics label $y_{data}$.
Correspondingly, the loss function denotes 
\begin{equation}
    C_4= [\,y_{data} - a_{\theta^\prime}(x)\,]^2 \;.
    \label{eq:C4}    
\end{equation}
In the adversarial training explained above, the constrainer network $a_{\theta^\prime}$ 
is trained supervised after each critic update.
As the loss $C_4$ will influence only the constrainer (critic and constrainer are separated networks), 
both networks could also be trained simultaneously.

A similar term as eq.~(\ref{eq:C3}) has been used in the so-called AC-GAN where
images were generated using label conditioning \cite{2016arXiv161009585O}.
Instead of the discrete classifier we use a continuous label here, along with the WGAN
concept.

\subsection{Generating signal patterns using an energy label}
\label{sec:1c}
Signals observed in particle detectors originate from physics-driven processes
which lead to patterns dissimilar from random patterns.
For example, low-energy events in a calorimeter typically exhibit signal patterns with small
signals and a small spatial extent, while high-energy events cause signal patterns that
can be widely distributed.

To enforce the generator to respect this dependency, we input an energy label $E_{label}$ in addition
to the random noise $z$ to generate a signal pattern for the detectors of our cosmic ray 
observatory. Therefore, the generator is modified to $g=g_{\theta}(z, E_{label})$.
The distribution of the input $E_{label}$ follows the energy distribution of the air shower simulation.
In this way, generated patterns are conditioned to follow the primary particle energy as reconstructed by the constrainer network. The resulting energy distribution of the generated events will cover a similar phase space as the simulated data.

\paragraph{Network architecture and training.}
\label{sec1:b2}
In order to generate signal patterns to a given energy label, our training architecture consists of a generator, a constrainer and a critic. The complete training architecture is shown in Fig.~\ref{fig:architecture_gan}. The generator and the critic networks are used for the adversarial training procedure explained above in 
section \ref{sec:1a}.

Our generator architecture is motivated by the class of DCGANs which is proposed in \cite{DCGAN} and is based 
on Transposed Convolutions. 
Its specification can be found in the Appendix Tab.~\ref{table:generator}. 

For the critic network we used an architecture inspired by \cite{2016arXiv161009585O} with 
LeakyReLU non-linearity and without batch normalization layers as we used the gradient penalty loss 
$C_2$ (\ref{eq:C2}). 
For details of the critic model see Tab.~\ref{table:critic}.

Also shown in Fig.~\ref{fig:architecture_gan} is the constrainer network which is constructed similarly to the architecture presented in \cite{Erdmann:2017str}. In the following we will refer to this architecture as AixNet.
It is used to reconstruct the energy contained in the signal patterns.
In our setup we used $l=80$ noise variables which are sampled from a Gaussian distribution. 
The loss weight $\kappa=0.001$ was used in eq. (\ref{eq:C3}), 
and the gradient penalty was scaled with $\lambda=3$ (eq.~(\ref{eq:C2})).

Furthermore, we used a batch size of $64$, and the training was run for $100$ epochs with $n_{\mathrm{critic}}=8$ critic and constrainer updates before $1$ generator update was applied. We used the Adam optimizer with $lr=0.0005$, $\beta_1=0.5$ and $\beta_2=0.9$ \cite{1412.6980}. Furthermore, a decay of $0.0001$ was used.
As deep learning framework we use Keras \cite{Keras} and TensorFlow \cite{tensorflow}. For training we used NVIDIA GeForce GTX 1080 cards provided by the VISPA project \cite{VISPA}.

\subsection{Refining simulated signal traces to match data}
\label{sec:refining}

Our second application of WGANs aims at refining simulated detector signal distributions to match data distributions.
This is an attempt to solve a long-standing issue in machine learning, namely 
training of deep networks with simulations that differ from data distributions.
Such refined simulations potentially increase the robustness of deep 
neural networks for data applications.

To refine signal traces we require the energy label of the simulation to follow a similar profile as the data distribution and make use of a generator network architecture which allows only for small modifications of the simulated traces.

\paragraph{Network architecture and training.}
Fig.~\ref{fig:architecture_refiner} shows the network architecture for refining
simulated signal distributions.
Here, the generator network of adversarial training is called `refiner' 
network \cite{2016arXiv161207828S}. On input it receives simulated signal distributions instead of 
random numbers and returns modified distributions with the same dimensionality as the original input.
The refiner network and the critic network are subjected to adversarial training as explained in
section~\ref{sec:1a}, where the refiner replaces the generator part.

In our example application of the cosmic ray observatory, for every event we simulate time traces for 
$d$ detectors placed on a hexagonal grid, each of which has $k$ time bins with amplitude $A_k$.
In total $d\times k$ amplitude values are given to the refiner network.
On output the refiner network again delivers $d\times k$ values as the modified
time traces for the $d$ detectors.

Correspondingly, the data pool contains time traces of data events.
These traces are unlabeled, i.e., the data traces have no direct relation with the generated time traces.

The refiner employs a residual architecture \cite{ResNet1} inspired by SimGAN \cite{2016arXiv161207828S} using 4 residual blocks, each consisting of two 3-dimensional convolutions with kernels operating on the time traces of the detector array. The architecture of the critic closely resembles the structure of AixNet \cite{Erdmann:2017str} later used to reconstruct the energy. The detailed network architecture for the refiner is listed in Tab.~\ref{table:refiner}.

The networks are trained for $10000$ refiner iterations with a batch size of $100$ following the algorithm outlined in section~\ref{sec:1a} using the distance measure as presented in (\ref{eq:C1}, \ref{eq:C2}). For each refiner step we update the critic $n_{\mathrm{critic}} = 10$ times with a gradient penalty scaled by $\lambda=5$ using the Adam optimizer \cite{1412.6980} with learning rate $lr=10^{-4}, \beta_1=0.5$ and $\beta_2=0.999$. We evaluate the final performance of the refining network by training AixNet \cite{Erdmann:2017str} on the refined traces to reconstruct the primary particle energy (see section~\ref{sec:refined_eval} below).

\section{Simulated data for training and evaluation}
\label{sec:3}

\begin{figure*}[t]
\captionsetup[subfigure]{aboveskip=-1pt,belowskip=-1pt}
\begin{centering}
\begin{subfigure}[b]{0.49\textwidth}
\includegraphics[trim={0 0 0 0},clip,,width=\textwidth]{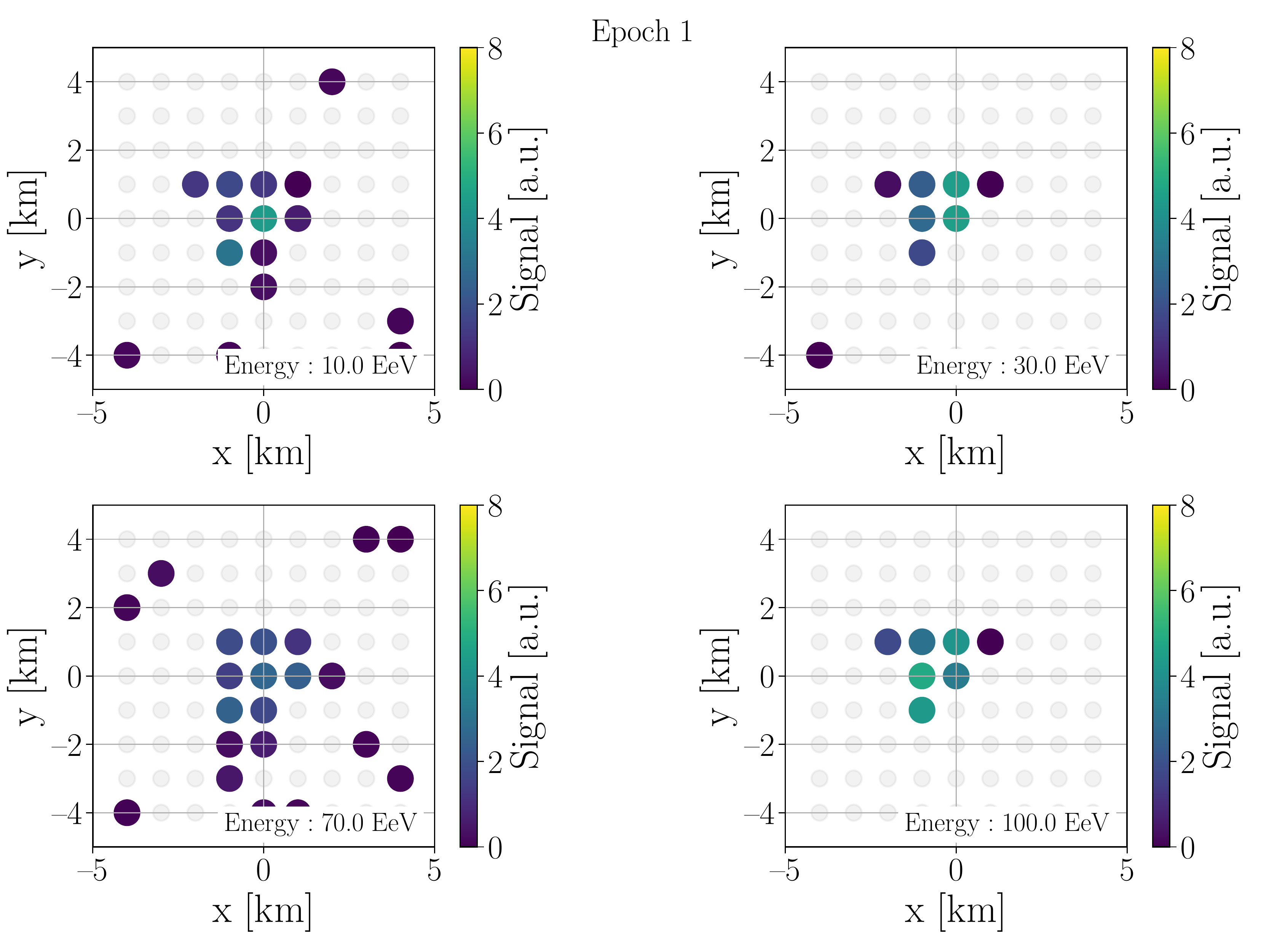}
\subcaption{}
\label{fig:maps_a}
\end{subfigure}
\hfill
\begin{subfigure}[b]{0.49\textwidth}
\includegraphics[trim={0 0 0 0},clip,,width=\textwidth]{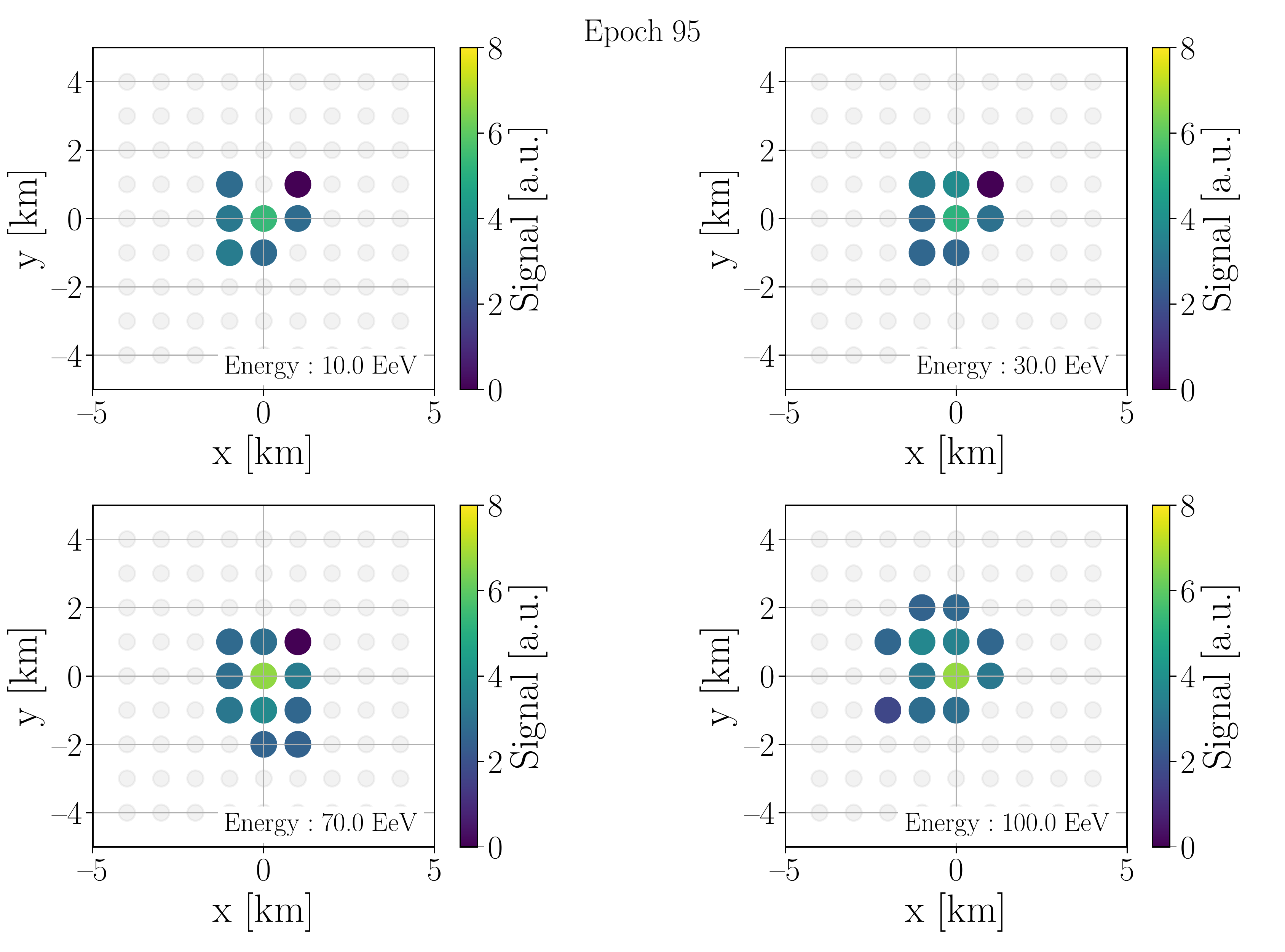}
\subcaption{}
\label{fig:maps_b}
\end{subfigure}
\caption{Detector signal patterns of WGAN generated events including an energy condition.
Color coded is the signal strength for four events with different cosmic ray energies after
(a) 1 training epoch,
(b) 95 training epochs.}
\label{fig:pattern}
\end{centering}
\end{figure*}

To simulate cosmic ray-induced air showers we use the parameterized simulation program
presented in \cite{Erdmann:2017str}.
This simulation directly produces signal traces in water-Cherenkov detectors placed on a
hexagonal grid with a spacing of $1500 \, \mathrm{m}$.
They are located at a height of $1400 \, \mathrm{m}$ above sea level, motivated by the Pierre Auger Observatory \cite{Auger-NIM-Pub}.
For simplicity we restrict our simulations to vertical showers with a fixed depth of the
shower maximum.
Alternatively, the setup can be understood as a granular calorimeter with a single readout layer.

For each simulated event, the air shower consists of two components, one component reflecting muons
from pion decays, the other being particles of the electromagnetic cascade which arrive
with a time delay compared to the muons.
The simulation has been tuned from measurements to deliver $\sim 30\%$ of the energy in
the muon component, and $\sim 70\%$ through the electromagnetic cascade.

\paragraph{Data.}
\label{sec:3a}
We simulated $10^5$ cosmic proton events with energies between $E=(1,...,100)$ EeV following a flat distribution ($1 \, \mathrm{EeV} = 10^{18} \, \mathrm{eV}$).
The muon and electromagnetic energies follow the above-mentioned $30/70$ subdivision.
Each event consists of $d=9\times 9$ detectors with signal traces containing $k=80$ amplitude
values in the time bins.

For each event, the time integrated signal strengths in the detectors can be visualized as a
two-dimensional signal pattern. Examples of signal patterns as well as of signal time traces will be presented in sections~\ref{sec:4} and \ref{sec:refined_eval} respectively.
We will refer this simulated data set to as our `data' events.

\paragraph{Simulation.}
\label{sec:3b}
In order to produce a simulation which deviates from the data, we produce another set
of $10^5$ simulated cosmic proton events with the same conditions as for the above-mentioned
data set, except for the division of the energy.
For the energy fractions of the muonic and electromagnetic energies we use the 
inverted $70/30$ subdivision instead. Furthermore, the amount of absolute noise in the time traces and event-by-event fluctuations are reduced by a factor of two in order to reflect underestimation of noise in detector simulations.

As the time of arrival and the transverse shower distributions are different
between muons and particles of the electromagnetic cascade, the shapes of the 
time traces are different compared to 
the traces of the data set, as shown in section~\ref{sec:refined_eval}.
We will refer to this set of simulated events as our `simulated' events.

\section{Energy constrained spatial signal pattern}
\label{sec:4}
To generate patterns of detector signals as the response to cosmic ray events
we use the network setup presented in Fig.~\ref{fig:architecture_gan}.
The events in the data pool originate from the data set described in section~\ref{sec:3a}, using only the $d=9\times 9$ values of the time integrated signal traces and the original energy of the primary particle.

In Fig.~\ref{fig:maps_a} we show example patterns of detector signals generated
after $1$ training epoch with test labels of $E_{label}= 10, 30, 70$ and $100$~EeV.
All patterns appear to be rather different from the typical patterns with large signals
in the shower center and smaller signals around that.
Furthermore, the sizes of the generated signal patterns are not in agreement with the energy labels.

In Fig.~\ref{fig:maps_b} we show example patterns after $95$ epochs, again with test labels of 
$E_{label}=$ $10$, $30$, $70$ and $100$~EeV.
Already here the signal patterns improve and are inline with our expectations. 
The hottest station is in the center of the shower and the signal decreases for outlying stations. 
The pattern structure also shows a highly local correlation of neighbor stations which coincide with expectations.
Furthermore, the increasing pattern size for higher energies is clearly visible. 
In addition, the total signal distribution correlates significantly with higher energies and meets with expectations.

To reconstruct the primary particle energy from the generated signal patterns we use
the simultaneously trained constrainer (see Fig.~\ref{fig:architecture_gan}).
\begin{figure}[b]
\includegraphics[trim={0.5cm 0 1cm 2cm},clip,,width=\linewidth]{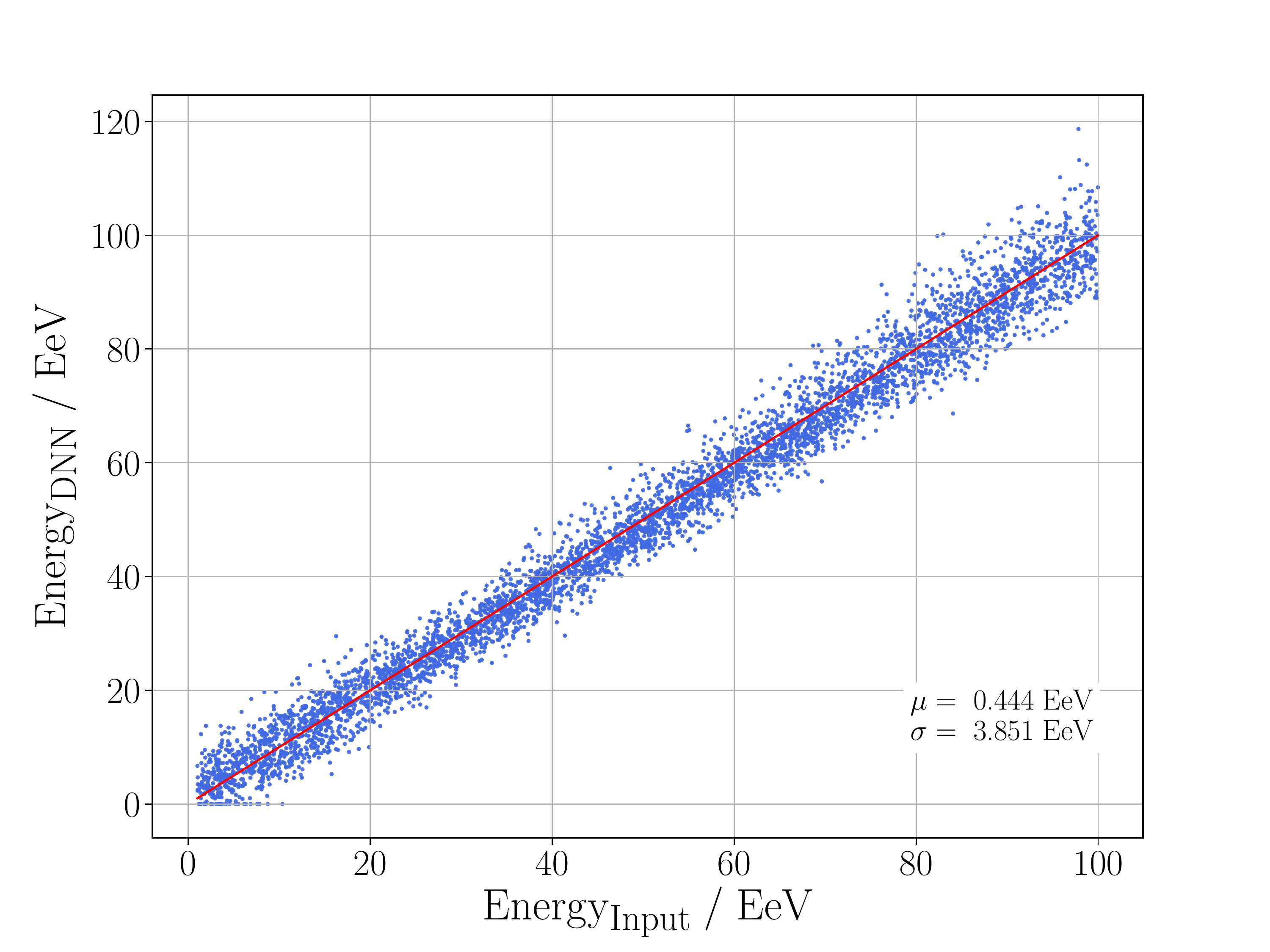}

\caption{Energy reconstructed by the constrainer network from analyzing the WGAN generated
signal patterns compared to the energy input to the WGAN.}
\label{fig:WGAN-energy}
\end{figure}

In Fig.~\ref{fig:WGAN-energy} we show the correlation of the input energy to the generator with the
particle energy as reconstructed by the constrainer network AixNet for $5000$ generated events.
The distinct correlation implies that the generator has not only been trained to produce detector signal
patterns as an image-like product, but has in addition learned to produce patterns related to a given particle energy.

\begin{figure}[h]
\begin{centering}
\begin{subfigure}[b]{\linewidth}
\includegraphics[trim={0.235cm 0 0 0.3cm},clip,,width=\textwidth]{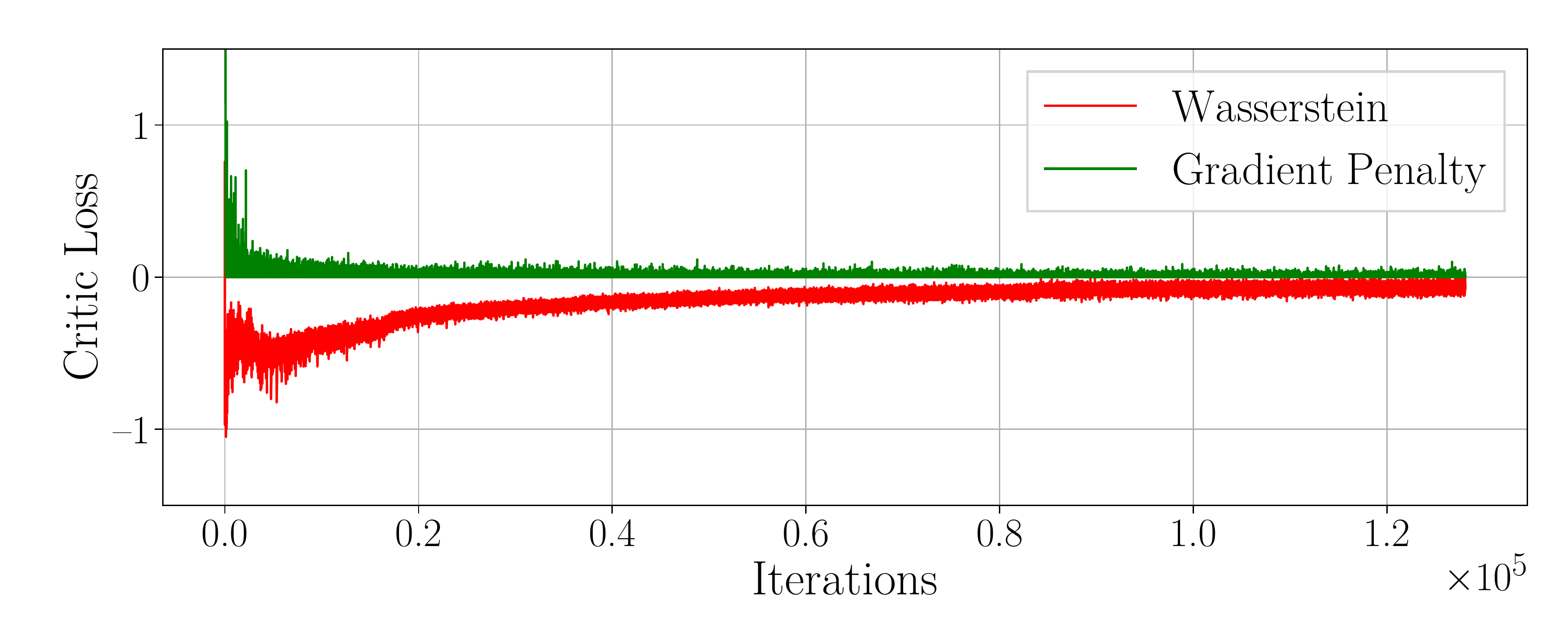}

\subcaption{}
\label{fig:WGAN-loss_a}
\end{subfigure}
\begin{subfigure}[b]{\linewidth}
\includegraphics[width=\textwidth]{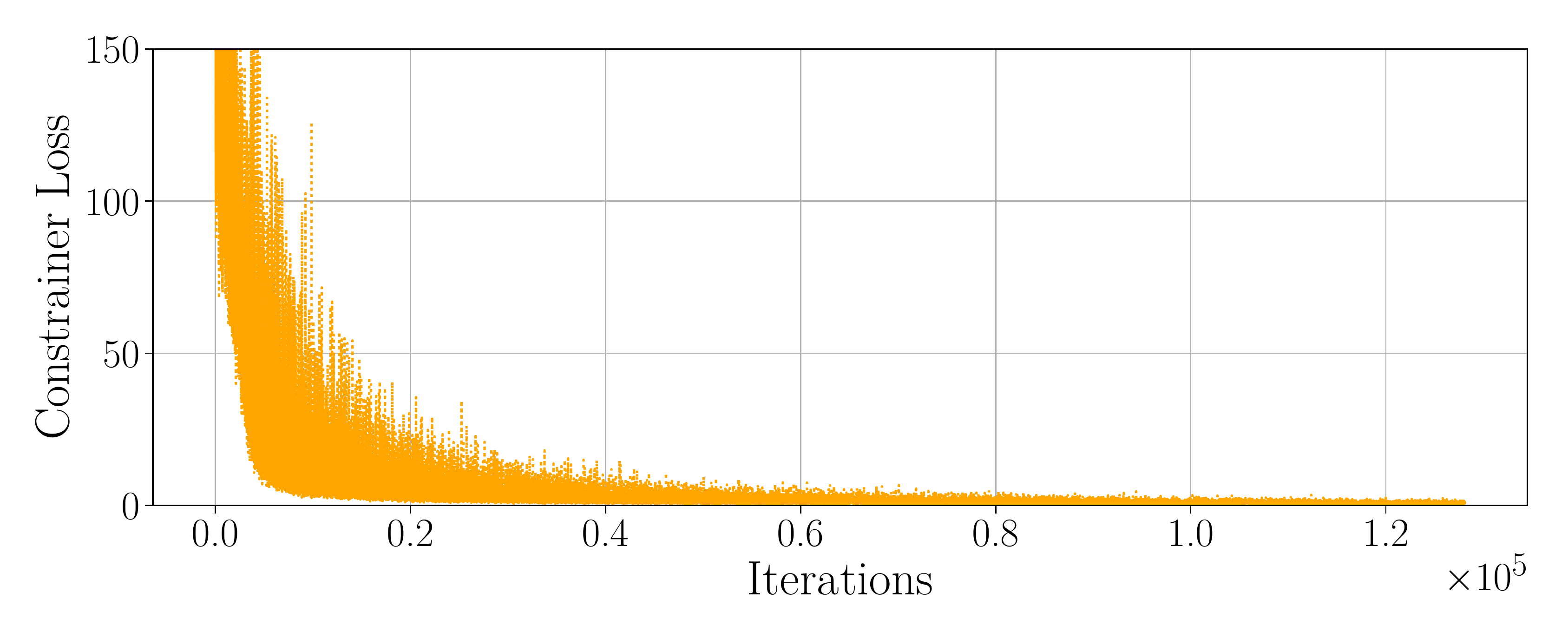}
\subcaption{}
\label{fig:WGAN-loss_b}
\end{subfigure}
\begin{subfigure}[b]{\linewidth}
\includegraphics[width=\textwidth]{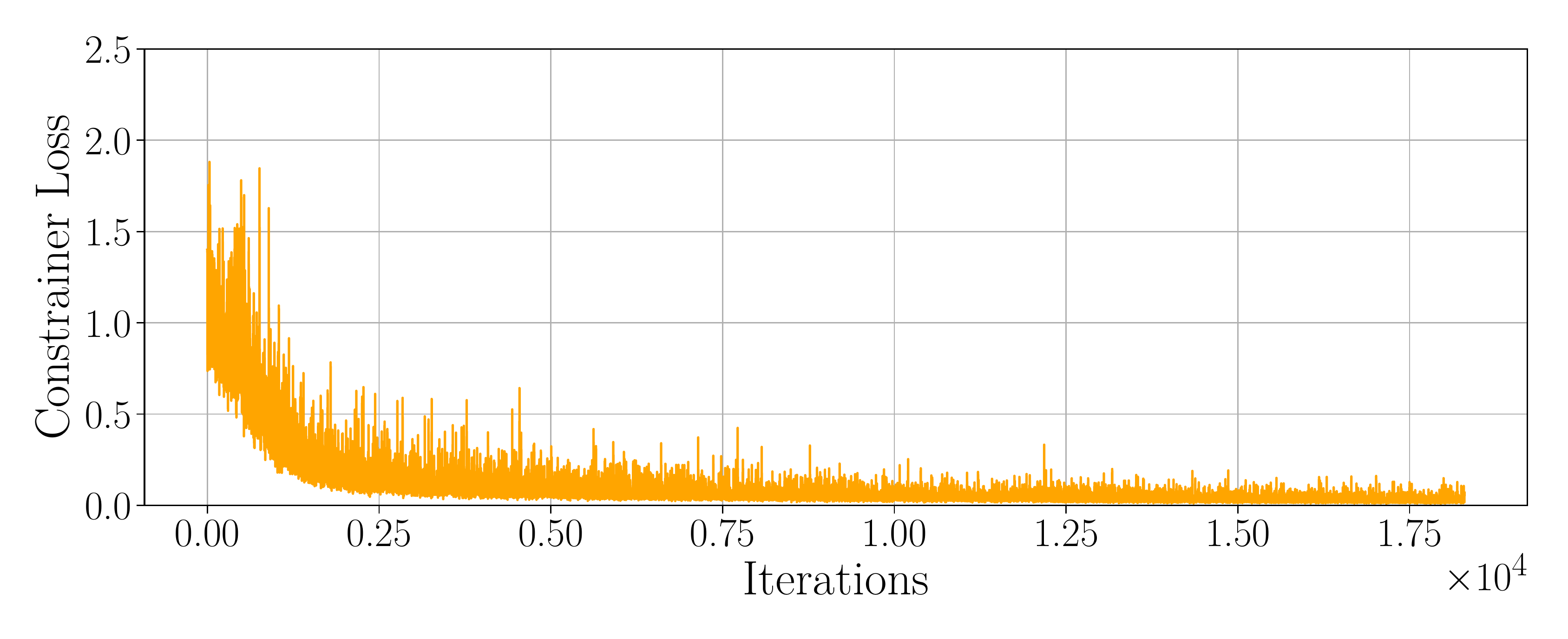}
\subcaption{}
\label{fig:WGAN-loss_c}
\end{subfigure}
\caption{Loss function (a) $-C_1$~(\ref{eq:C1}) during the critic training showing 
the approximated Wasserstein distance (red) between generated and simulated signal 
patterns, and the gradient penalty $C_2$ (\ref{eq:C2}) (green) scaled by $\lambda$, 
(b) $C_4$ (\ref{eq:C4}) of constrainer during the supervised training, and 
(c) $C_3$ (\ref{eq:C3}) of the constrainer scaled by the hyperparameter $\kappa$ during 
the generator training visualizing the increasing conditioning performance. Note that the constrainer is only updated every $n_{\mathrm{critic}}=8$ iterations.}
\label{fig:WGAN-loss}
\end{centering}
\end{figure}

In Fig.~\ref{fig:WGAN-loss_a} we show the critic loss of the adversarial training. 
The approximate Wasserstein distance $C_1$ (\ref{eq:C1}) (red curve) converges slowly 
to zero for increasing epochs. 
Furthermore, the gradient penalty $C_2$ (\ref{eq:C2}) (green curve) is reduced during 
the iterations. 
Hence the loss $C_1$ gives an estimation of the Wasserstein distance and therefore a similarity 
measure of the generated and the data events. 
The convergence to zero is in accordance with the generated events to reproduce expected 
properties (Fig.~\ref{fig:pattern}).

In Fig.~\ref{fig:WGAN-loss_b} the supervised training loss $C_4$ (\ref{eq:C4})
of the constrainer network AixNet reflects the improving reconstruction performance. 
We checked the validation loss as well (not shown in the figure) which shows no sign of overtraining.

Fig.~\ref{fig:WGAN-loss_c} shows the loss function $C_3$ (\ref{eq:C3}) during the generator 
training. 
The constrainer loss decreases considerably with increasing iterations. 
This development is also visible in Fig.~\ref{fig:pattern} where a correlation 
between signal pattern, signal size, and energy is apparant only for later epochs.

\section{Network training with refined signal traces}
\label{sec:refined_eval}
To reconstruct the primary particle energy from detector signals we 
will make direct use of the amplitude distributions of the time traces.
We again perform the energy reconstruction with AixNet \cite{Erdmann:2017str}.

Usually the training of a network is based on simulated data.
However, when reconstructing particle energies from measured traces, 
differences between data and simulated traces may cause substantial 
uncertainties in the reconstructed energy.
In order to reduce these uncertainties we will refine simulated traces to match unlabeled 
data-like traces using the adversarial network architecture presented in Fig.~\ref{fig:architecture_refiner}.
The refined traces will then be used to train AixNet.

\begin{figure}[b]
\begin{centering}
\begin{subfigure}[b]{\linewidth}
\includegraphics[width=\textwidth]{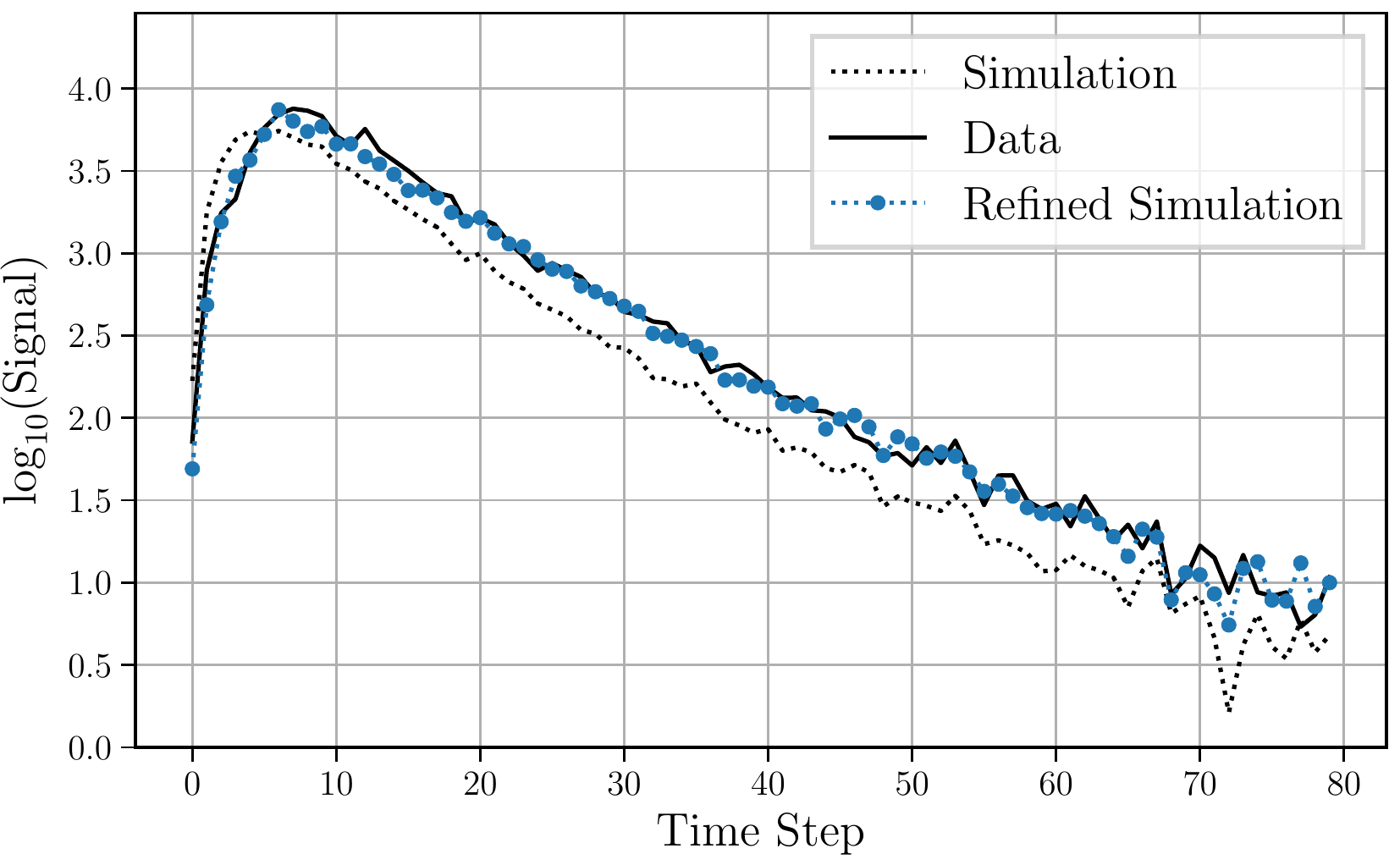}
\subcaption{}
\label{fig:traces_a}
\end{subfigure}
\begin{subfigure}[b]{\linewidth}
\includegraphics[width=\textwidth]{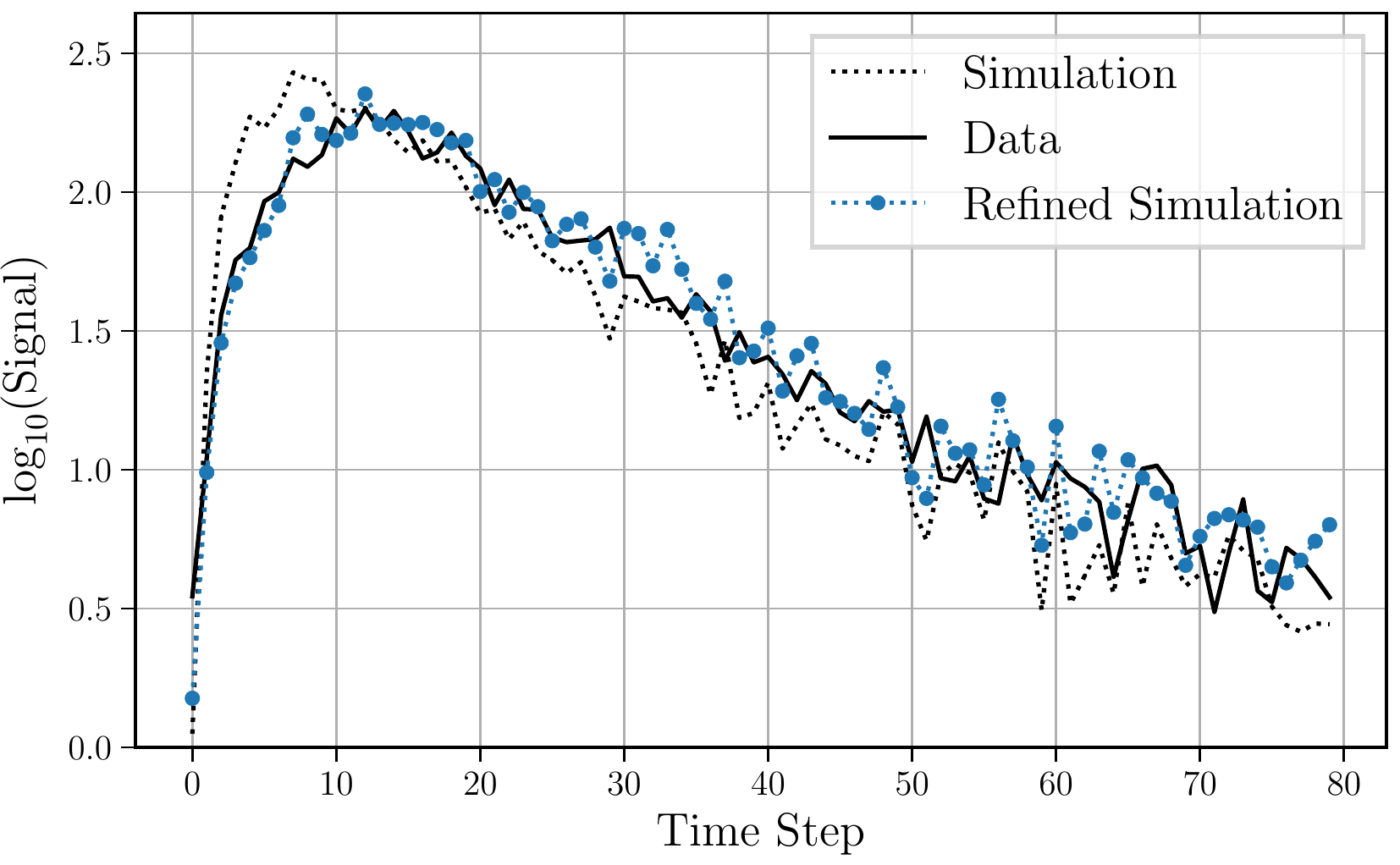}
\subcaption{}
\label{fig:traces_b}
\end{subfigure}
\caption{Signal traces of an event with energy $E=69\; \mathrm{EeV}$ 
measured by (a) the detector with the largest signal, 
(b) a neighbor detector. 
The black solid curve shows the data trace, the black dotted curve the trace from simulation, and
the blue circular symbols represent the refined simulated traces.
}
\label{fig:traces}
\end{centering}
\end{figure}

In Fig.~\ref{fig:traces_a} we show the time trace of the detector with the largest signal in a data 
event with particle energy $E=69\; \mathrm{EeV}$ by the black solid curve.
The black dotted curve represents a corresponding simulated time trace with matching primary energy.
Due to the overestimated muon component in the simulated shower, the amplitude of the simulated time trace 
rises faster than in the corresponding data event (for a detailed definition of the data sets refer to section~\ref{sec:3a}).
As shown by the blue circular symbols in Fig.~\ref{fig:traces_a},
the refiner network modified the simulated trace to more closely resemble the data trace.
\begin{figure*}[t]
\begin{centering}
\begin{subfigure}[b]{0.32\textwidth}
\includegraphics[width=\textwidth]{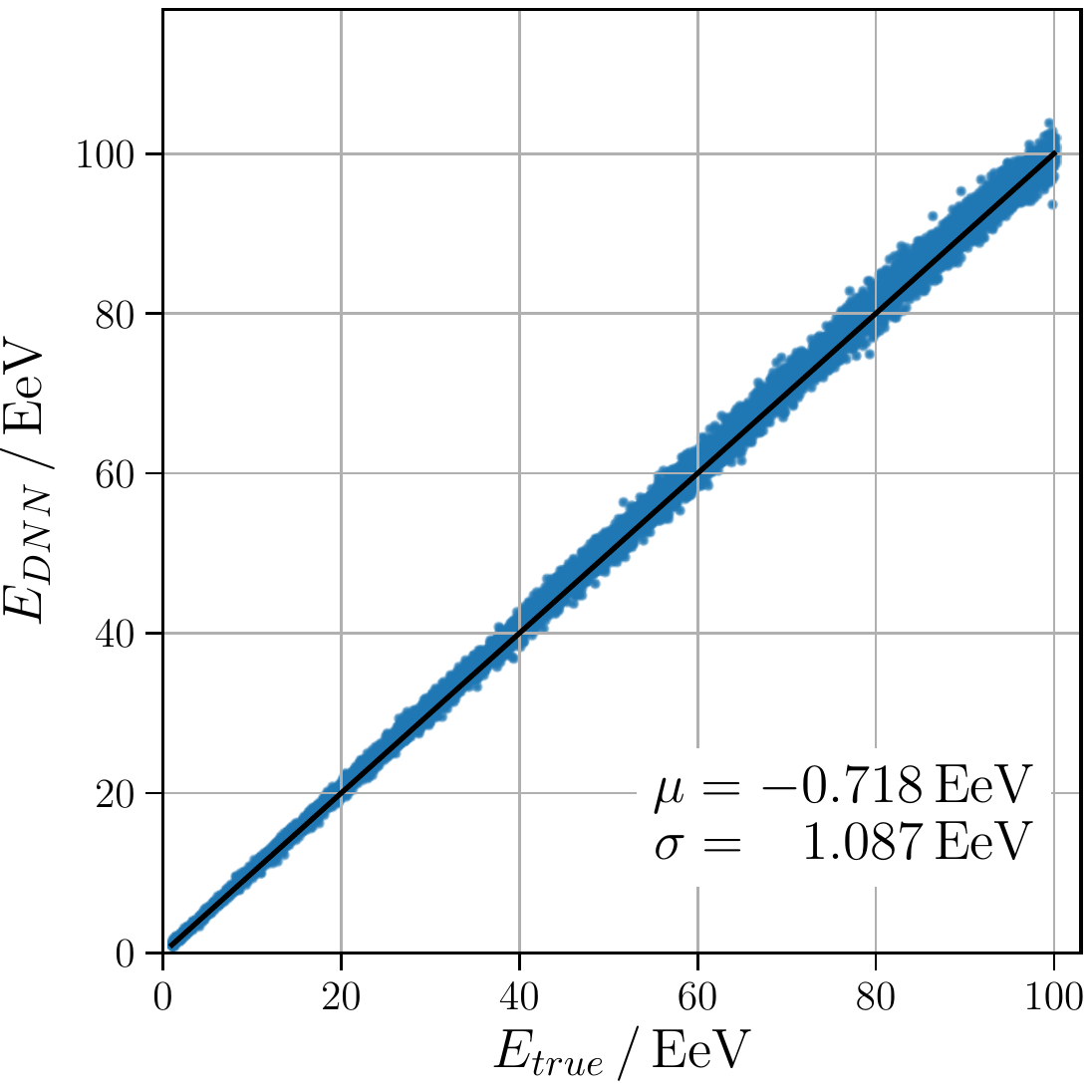}
\subcaption{}
\label{fig:REFINE-energy_a}
\end{subfigure}
\hfill
\begin{subfigure}[b]{0.32\textwidth}
\includegraphics[width=\textwidth]{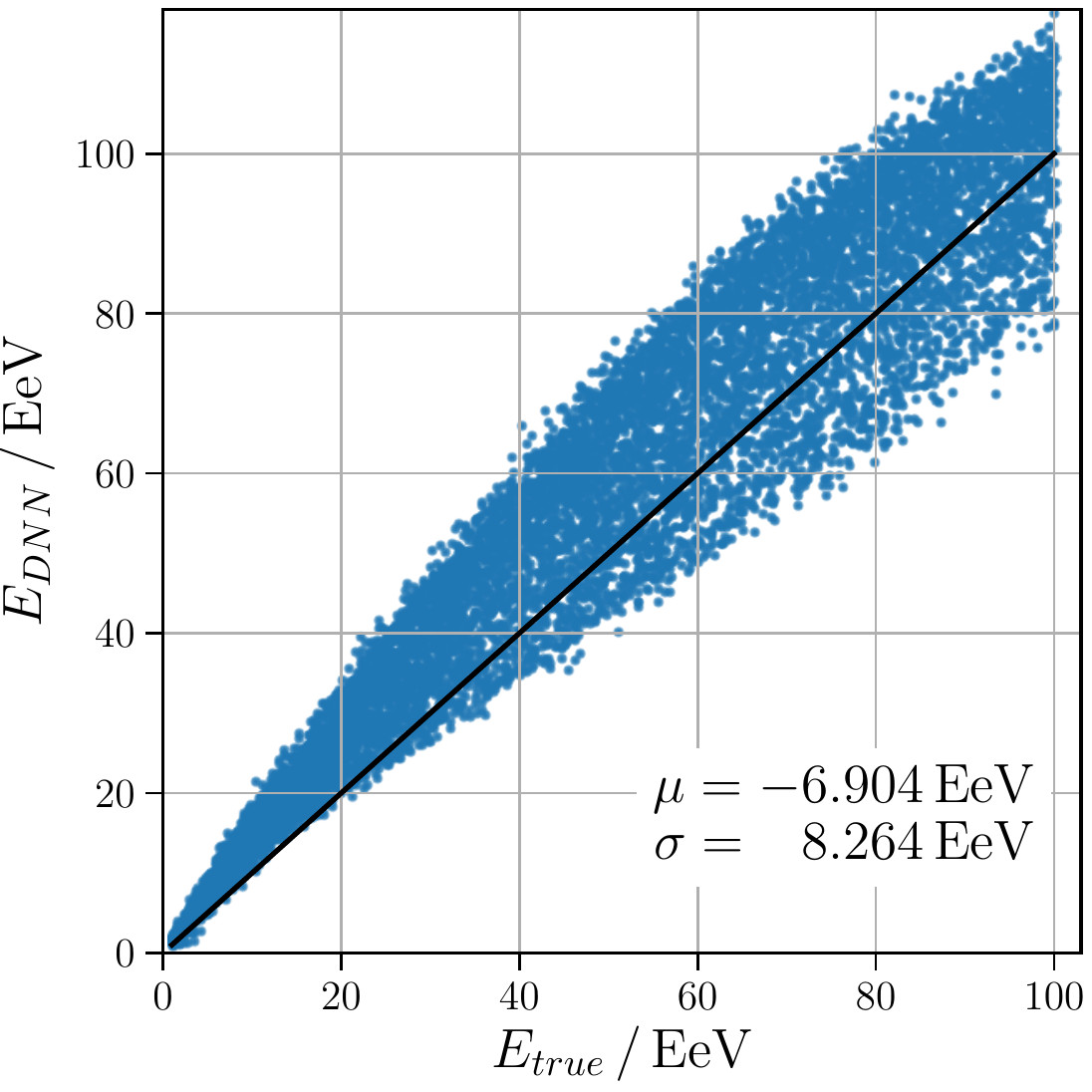}
\subcaption{}
\label{fig:REFINE-energy_b}
\end{subfigure}
\hfill
\begin{subfigure}[b]{0.32\textwidth}
\includegraphics[width=\textwidth]{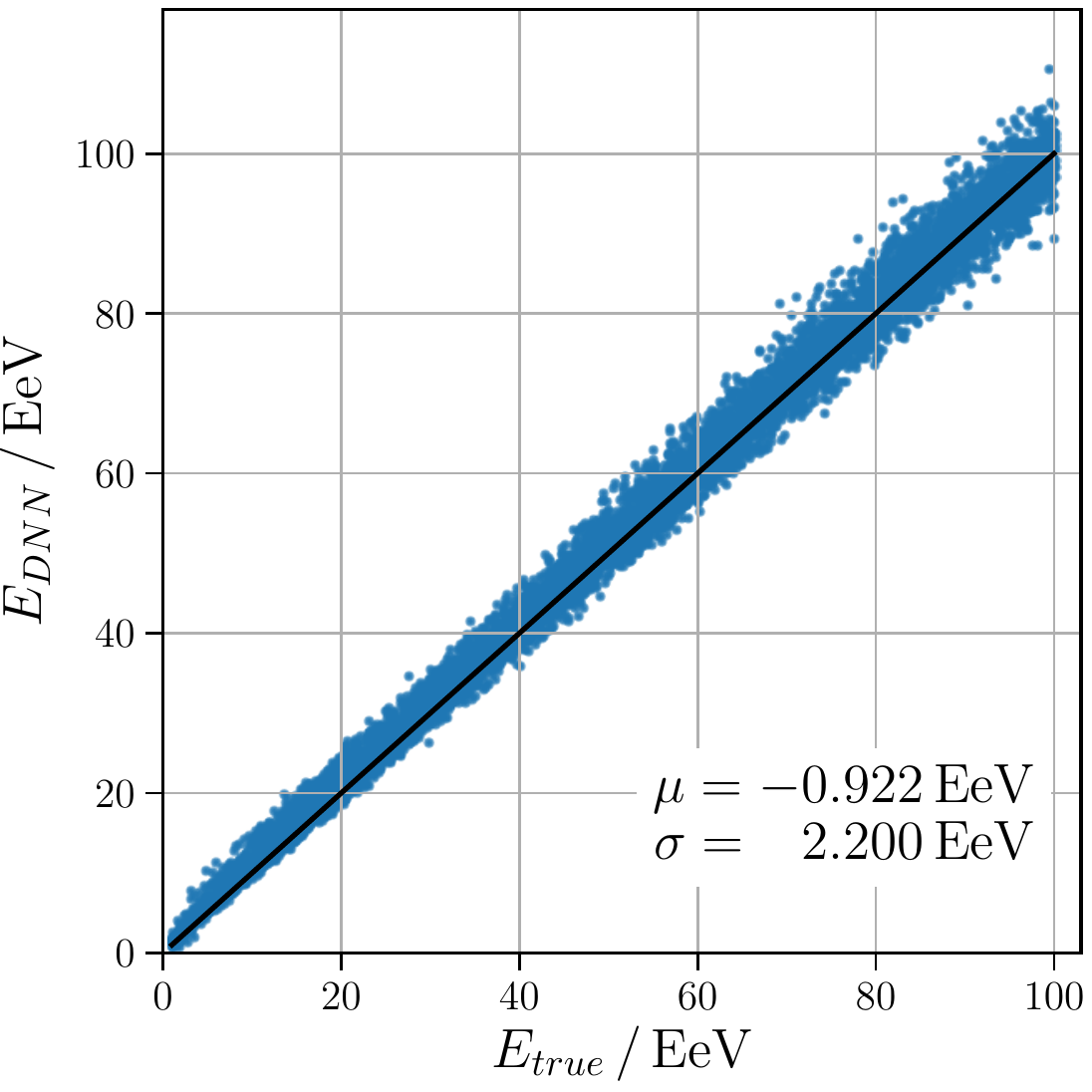}
\subcaption{}
\label{fig:REFINE-energy_c}
\end{subfigure}
\caption{Reconstruction of the primary particle energy using the signal traces
of the detectors as input to AixNet.
(a) Benchmark for training and evaluation, both based on simulated data.
(b) Reconstructed data events using the network of (a) trained on simulated data. 
(c) Reconstructed data events with AixNet trained by simulated traces that
were refined to match data events prior to the training (compare Fig.~\ref{fig:traces}).}
\label{fig:REFINE-energy}
\end{centering}
\end{figure*}
Note that this direct comparison between data traces and simulated traces results from a test data set simulated with identical parameters and identical random seeds.
For the unsupervised training of the networks, this matched information is not available as the data 
traces are simulated with different random seeds and are passed unlabeled to the critic network.

In Fig.~\ref{fig:traces_b} we show the signal traces for a neighbor detector.
Also here, the originally simulated trace is adapted by the refiner network to match the data trace.

To evaluate the ability to preserve the properties of the simulation we investigate the 
impact of refined traces on the energy reconstruction. 
We trained AixNet to reconstruct the primary particle energy on the originally simulated tra\-ces, or alternatively the refined simulated tra\-ces.

In Fig.~\ref{fig:REFINE-energy_a},~\ref{fig:REFINE-energy_b} we trained AixNet on the originally 
simulated traces.
In Fig.~\ref{fig:REFINE-energy_a} we benchmark AixNet by reconstructing the particle energy on a test 
set of simulated traces following the same distribution as the simulated training data. 
This demonstrates a good energy reconstruction quality of the network.

In Fig.~\ref{fig:REFINE-energy_b} we reconstruct particle energies of data events with the previous 
network trained on simulated traces. 
The network generalizes poorly on data due to the dissimilarities of the training set 
(simulated) and test set (data) which leads to a non-linear reconstruction bias and
increased reconstruction uncertainties. 
This is a common problem when training neural networks on simulations that do not 
perfectly mirror real data.

In Fig.~\ref{fig:REFINE-energy_c} we trained AixNet on the refined 
simulated traces instead of the original simulated traces and again evaluated 
the network performance on data traces. 
The network performs remarkably better compared to the training with the originally simulated data. 
The reconstruction quality is found to be worse compared to the benchmark shown in Fig.~\ref{fig:REFINE-energy_a}. 
However, compared to the training with the original simulation (Fig.~\ref{fig:REFINE-energy_b}),
training with refined traces leads to a lower energy bias and improved energy resolution
This shows that the refiner network is able to modify simulations to more accurately resemble the data distribution.

\begin{figure}[h]
  \includegraphics[width=\linewidth]{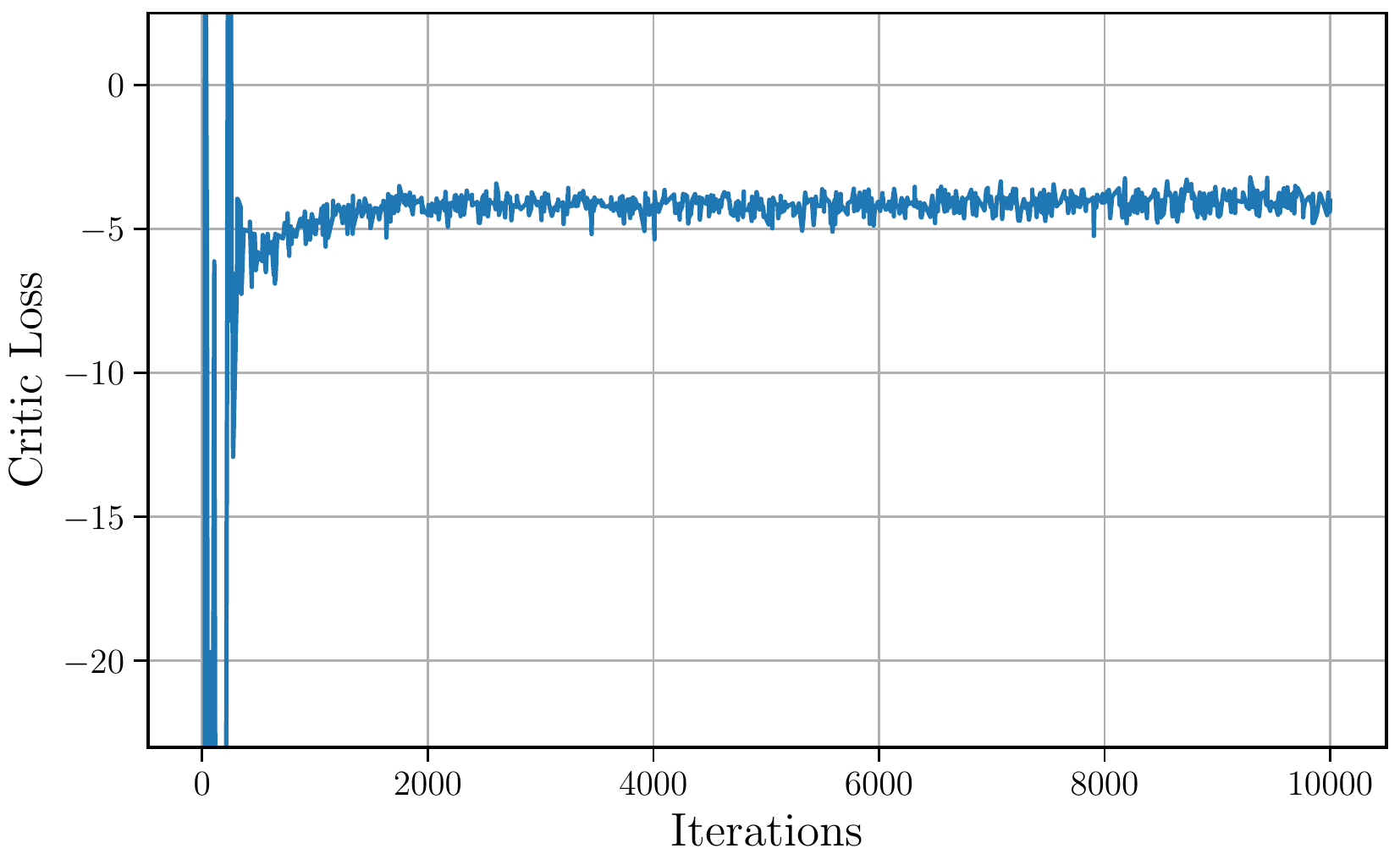}
\caption{Loss function of the critic network reflecting the similarity of the simulated traces
and data traces.}
\label{fig:REFINE-loss}
\end{figure}

In Fig.~\ref{fig:REFINE-loss} we show the convergence of the critic loss $-C_1 + C_2$ (\ref{eq:C1},
\ref{eq:C2}) as a similarity measure of the refined and the data traces. 
With an increasing number of iterations, the refiner network is able to adapt simulations 
to better resemble data.
The converged distance measure indicates remaining differences between data and simulation.
However, the impact of these differences appear to be sufficiently small when evaluating 
the quality of the energy reconstruction.

\section{Conclusion}

In this paper we investigated two variants of adversarial network methods for
detector simulations.
In both cases, the transfer of probability distributions from one data set to
another data set by unsupervised network training is found to work well 
using the Wasserstein distance in the loss function.
As a specific example we used air shower simulations with an array of ground-based
water-Cherenkov detectors to represent single-layer calorimeter
simulations.

We generated signal patterns of detector responses
showing that the patterns can be constrained to follow properties expected 
from physics.
In our example we constrained the energy contained in the shower and found that
the generated events follow this given energy.

Refinement of simulated detector signal traces to match data traces appears 
to be a promising method for solving a long-standing issue in machine learning.
Instead of training a deep network with simulations that differ in details 
from data, simulations can be adapted to match data prior to network training.
For our example of the air shower simulations we showed that small refinements
of the signal traces lead to improved reconstruction of the primary particle 
energy with respect to both energy bias and energy resolution.

\begin{acknowledgements}
This work is supported by the Ministry of Innovation, Science and Research of the State of North Rhine-Westphalia, and the Federal Ministry of Education and Research (BMBF). We wish to thank Thorben Quast for his valuable comments on the manuscript.
\end{acknowledgements}

\begin{conflicts}
On behalf of all authors, the corresponding author
states that there is no conflict of interest.
\end{conflicts}

\eject
\appendix
\onecolumn
\section{Appendix}
\begin{table}[!h]
\centering
\caption{Generator network as used in the WGAN to generate signal patterns.}
\label{table:generator}
 \begin{tabular}{c c c c c c}
 \toprule
 Operation & Kernel & Feature Maps & Padding & BN & Activation \\ \midrule
 \multicolumn{6}{c}{Generator $80+1$ Input} \\
 \midrule
 Linear & N/A & 80 & & $\times$ & ReLU\\ 
 Transposed Convolution & $3\times$3 & 64 & valid & $\surd$ & ReLU\\
 Transposed Convolution & $3\times$3 & 128 & valid & $\surd$ & ReLU\\
 Transposed Convolution & $3\times$3 & 128 & valid & $\surd$ & ReLU\\
 Transposed Convolution & $3\times$3 & 256 & valid & $\surd$ & ReLU\\
 Convolution & $3\times$3 & 1 & same & $\times$ & ReLU\\
 \midrule
 \multicolumn{6}{c}{Generator $9\times 9\times 1$ Output} \\
 \bottomrule
\end{tabular}
\end{table}

\begin{table}[!h]
\centering
\caption{Critic network as used in the WGAN to generate signal patterns.}
\label{table:critic}
 \begin{tabular}{c c c c c c} 
 \toprule
 Operation & Kernel & Feature Maps & Padding & BN & Activation \\ \midrule
 \multicolumn{6}{c}{Critic $9\times9\times1$ Input}\\ \midrule
 Convolution & $3\times$3 & 64 & same & $\times$ & LeakyReLU\\
 Convolution & $3\times$3 & 128 & same & $\times$ & LeakyReLU\\
 Convolution & $3\times$3 & 128 & same & $\times$ & LeakyReLU\\
 Convolution & $3\times$3 & 256 & same & $\times$ & LeakyReLU\\
 GlobalMaxPooling & & & & $\times$ &\\
 Dropout & & & & &\\
 Linear & N/A & 100 & & $\times$ & LeakyReLU\\
 Dropout & & & & &\\
 Linear & N/A & 1 & & $\times$ & \\ \midrule
 \multicolumn{6}{c}{Critic $1$ Output}\\ \bottomrule
\end{tabular}
\end{table}

\begin{table}[!h]
\centering
\caption{Refiner network as used in the WGAN to refine signal traces.}
\label{table:refiner}
\begin{tabular}{cccccc}\toprule
Merge Operation & Operation	& Kernel & Feature Maps & Padding & Activation \\ \midrule
\multicolumn{6}{c}{$9\times 9\times 80 \times 1$ Input} \\ \midrule
\multirow{2}{*}{Addition} & Convolution & $1\times 1 \times 7$ & $64$ & same & ReLU \\
&Convolution & $1\times 1 \times 7$ & $64$ & same & ReLU \\ \midrule
\multirow{2}{*}{Addition} &Convolution & $1\times 1 \times 7$ & $64$ & same & ReLU \\
&Convolution & $1\times 1 \times 7$ & $64$ & same & ReLU \\ \midrule
\multirow{2}{*}{Addition} &Convolution & $1\times 1 \times 7$ & $64$ & same & ReLU \\
&Convolution & $1\times 1 \times 7$ & $64$ & same & ReLU \\ \midrule
\multirow{2}{*}{Addition} &Convolution & $1\times 1 \times 7$ & $64$ & same & ReLU \\
&Convolution & $1\times 1 \times 7$ & $64$ & same & ReLU \\ \midrule
&Convolution & $1\times 1 \times 1$ & $1$ & same & ReLU \\ \midrule
\multicolumn{6}{c}{$9\times 9\times 80 \times 1$ Output} \\ \bottomrule
\end{tabular}
\end{table}

\end{document}